\documentclass{appolb}
\usepackage{graphicx}
\begin{document}
\title{Accessing transversity GPDs in neutrino-production of a charmed meson%
\thanks{Presented at 16th conference on Elastic and diffractive scattering, EDS Blois 2015, Borgo, Corsica, France, June 29th-July 4th, 2015}%
}
\author{B. Pire
\address{Centre de physique th\'eorique, \'Ecole Polytechnique,
CNRS, F-91128 Palaiseau}
\vspace*{.3cm}
\\
{L. Szymanowski
}
\address{National Centre for Nuclear Research (NCBJ), 00-681 Warsaw, Poland}
}
\maketitle
\begin{abstract}
We demonstrate that the transversity chiral odd generalized parton distributions  can be accessed  through the azimuthal dependence of the $\nu_l N \to l^- D^+ N'$ or $\bar\nu_l N \to l^+ D^- N'$ differential cross sections, in the framework of the colinear QCD approach, where GPDs factorize from  perturbatively calculable coefficient functions calculated up to  order $m_c/Q$. 
\end{abstract}
  
\section{Introduction}
Thanks to quark mass effects, flavor changing electroweak interactions open new ways to access chiral-odd quantities like the transversity generalized parton distributions (GPDs) in a nucleon  \cite{Pire:2015iza}.
The transversity distributions which encode the  transverse spin structure of the nucleon have proven to be among  the most difficult hadronic quantities to access. This is  due to the chiral odd character of the quark operators which enter their definition; this feature enforces the decoupling of these distributions from  most measurable hard amplitudes. After the pioneering works \cite{trans}, much effort \cite{Barone} has been devoted to the exploration of many channels but experimental difficulties have challenged some of the most promising ones. 
 
Generalized parton distributions  give access to the internal structure of hadrons in a much more detailed way than parton distributions measured in inclusive processes, since they allow a 3-dimensional analysis \cite{3d}. The study of exclusive reactions mediated by a highly virtual photon in the generalized Bjorken regime benefits from the factorization properties of the leading twist QCD amplitudes \cite{fact1,fact2,fact3}
of reactions such as deeply virtual Compton scattering. 

Neutrino production offers another way to access (generalized) parton distributions  \cite{weakGPD}. Although neutrino induced cross sections are orders of magnitudes smaller than those for electroproduction, the advent of new generations of neutrino experiments will open new possibilities. We want here to stress that they can help to access the elusive \cite{notrans} chiral-odd generalized parton distributions.

\section{The processes}
We consider \cite{Pire:2015iza} the exclusive reactions
\begin{eqnarray}
\nu_l (k)N(p_1) &\to& l^- (k')D^+ (p_D)N'(p_2) \,,\\
 \bar\nu_l (k) N(p_1) &\to& l^+ (k') D^-(p_D) N' (p_2)\,,\nonumber
\end{eqnarray}
in the kinematical domain where collinear factorization  leads to a description of the scattering amplitude 
in terms of nucleon GPDs and the $D-$meson distribution amplitude, with the hard subprocess  ($q=k'-k; Q^2 = -q^2$):
\begin{equation}
W^+(q) d \to D^+ d' ~~~~~\mbox{or} ~~~ W^-(q) u \to  D^- u'\,,
\end{equation}
described for the neutrino case by the  handbag Feynman diagrams of Fig. 1.
We use the standard notations of deep exclusive leptoproduction, namely $P=(p_1+p_2)/2$, $\Delta = p_2-p_1$, $t=\Delta^2$,  $x_B=Q^2/2p_1.q$, $y=p_1.q/p_1.k$ and $\epsilon \simeq 2(1-y)/[1+(1-y)^2]$. $p$ and $n$ are light-cone vectors ($v.n=v^+, v.p=p^-$ for any vector $v$) and $\xi=-\Delta.n/2P.n$ is the skewness variable. 

%%%%%%%%%%%%%%%%%%%%%%%%%%%%%%%%%%%%%%%%%%%%%%%%%%%%%
\begin{figure}
\includegraphics[width=12cm]{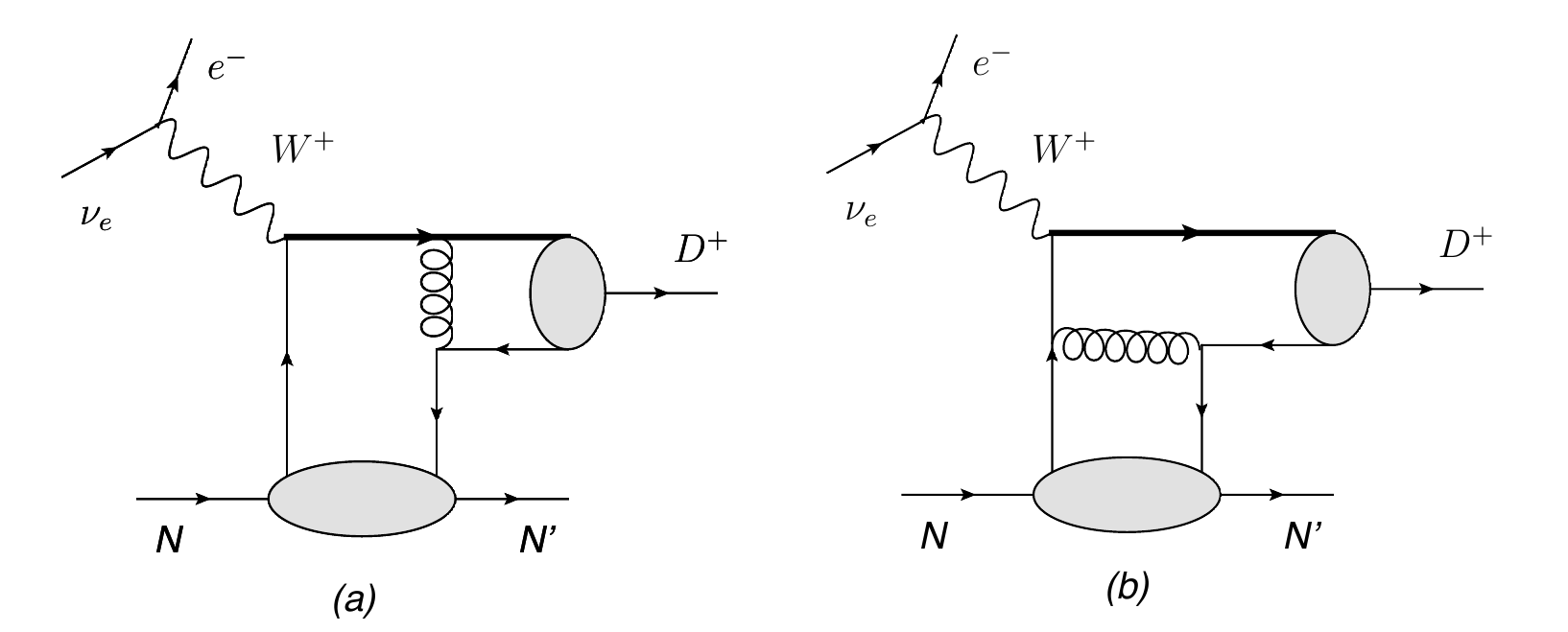}
\vspace{-.3cm}
\caption{Feynman diagrams for the factorized  amplitude for the $ \nu_e N \to \e^-  D^+ N'$ process; the thick line represents the heavy quark. In the Feynman gauge, diagram (a) involves convolution with both the transversity GPDs and the chiral even ones, whereas diagram (b) involves only chiral even GPDs.}
   \label{Fig1}
\end{figure}

 \subsection{Transversity GPDs.}

The four  twist 2 quark transversity GPDs have been defined \cite{transGPDdef} as : 
\begin{eqnarray}
  \label{tGPD}
\lefteqn{ \frac{1}{2} \int \frac{d z^-}{2\pi}\, e^{ix P^+ z^-}
  \langle p_2,\lambda'|\, 
     \bar{\psi}(-{\textstyle\frac{1}{2}}z)\, i \sigma^{+i}\, 
     \psi({\textstyle\frac{1}{2}}z)\, 
  \,|p_1,\lambda \rangle \Big|_{z^+= \mathbf{z}_T=0} } 
\nonumber \\
&=& \frac{1}{2P^+} \bar{u}(p_2,\lambda') \left[
 H_T^q\, i \sigma^{+i} +
 \tilde H_T^q\, \frac{P^+ \Delta^i - \Delta^+ P^i}{m_N^2} \right.
\nonumber \\
&& \left. +
  E_T^q\, \frac{\gamma^+ \Delta^i - \Delta^+ \gamma^i}{2m_N} +
 \tilde E_T^q\, \frac{\gamma^+ P^i - P^+ \gamma^i}{m_N}
  \right] u(p_1,\lambda).
\hspace{2em}
\end{eqnarray}
and their experimental access is much discussed in \cite{transGPDacc}. 
The leading GPD $H_T(x,\xi,t)$ is equal to the transversity PDF in the $\xi=0 ~, t=0$ limit. Models have been proposed \cite{transGPDacc,models} and some  lattice calculation results exist \cite{lattice}.

\subsection{The transverse amplitude up to $O(m_c/Q^2)$.}

It has been demonstrated \cite{Collins:1998rz} that hard-scattering factorization of meson leptoproduction  is valid at leading twist with the inclusion of heavy quark masses in the hard amplitude.  In our case this means including the part $\frac{m_c}{k_c^2-m_c^2}$ in the off-shell heavy quark propagator (see the Feynman graph  on Fig. 1a) present in the leading twist coefficient function. 
We keep the term  $m_c^2$ in the denominator since it will help us to understand precisely how to perform the integration over the longitudinal momentum fraction $x$ around the point $x= \xi$. 
The longitudinal leading twist amplitude with a longitudinally polarized $W$ boson $T_L$ has been computed in \cite{Kopeliovich:2012dr}. Adding mass terms to the heavy quark propagator  has no effect on the calculation of this amplitude,  but leads to a non-zero transverse amplitude when a chiral-odd transversity GPD is involved.

In the Feynman gauge, the non-vanishing $m_c-$dependent part of the Dirac trace in the hard scattering part depicted in Fig. 1a reads:
\begin{eqnarray}
Tr&[&\sigma^{pi}\gamma^\nu \hat p_D \gamma^5 \gamma_\nu \frac{m_c}{k_c^2-m_c^2+ i \epsilon}(1- \gamma^5)\hat \epsilon \frac{1}{k_g^2+ i \epsilon} ] \\
&& = \frac{2 Q^2}{\xi} \epsilon_\mu[i \epsilon^{\mu p i n} - g^{\mu i}_\perp]  \frac{m_c}{k_c^2-m_c^2+ i \epsilon} \frac{1}{k_g^2 + i \epsilon} \, , \nonumber
 \end{eqnarray} 
 where $k_c$ ($k_g$) is the heavy quark (gluon) momentum and $\epsilon$ the polarization vector of the $W-$boson (we denote $\hat p = p_\mu \gamma^\mu$ for any vector $p$; the transverse index $i$ will be contracted with the analogous index in transversity GPDs).
The fermionic trace vanishes for the diagram shown on Fig. 1b. The transverse amplitude is then written  as ($\tau = 1-i2$):
\begin{eqnarray}
T_{T} & = &\frac{iC \xi m_c}{\sqrt 2 Q^2}  \bar{N}(p_{2}) \left[  {\mathcal{H}}_{T}^\phi i\sigma^{n\tau} +\tilde {\mathcal{H}}_{T}^\phi \frac{\Delta^{\tau}}{m_N^2} \right. %\nonumber \\
%&&  
+ {\mathcal E}_{T}^\phi \frac{\hat n \Delta ^{\tau}+2\xi  \gamma ^{\tau}}{2m_N} - \tilde {\mathcal E}_{T}^\phi \frac{\gamma ^{\tau}}{m_N}] N(p_{1}), \nonumber
\end{eqnarray}
in terms of  transverse form factors that we define as  :
\begin{eqnarray}
{\cal F }_T^\phi=f_{D}\int \frac{\phi(z)dz}{\bar z}\hspace{-.1cm}\int \frac{F^d_T(x,\xi,t) dx }{(x-\xi+i\epsilon) (x-\xi +\alpha \bar z+i\epsilon)},
\label{TFF}
 \end{eqnarray} 
where $F^d_T$ is any d-quark transversity GPD, $\alpha = \frac {2 \xi m_c^2}{Q^2+m_c^2}$ and  ${\bar {\mathcal{E}}_T^\phi}=\xi{\mathcal{E}}_T^\phi-{\tilde {\mathcal{E}}_T^\phi}$. Keeping $\alpha$ in the denominator in Eq. (\ref{TFF}) regulates the treatment of the double pole in the integration over $x$.

\section{The angular dependence}
The dependence of a leptoproduction  cross section on azimuthal angles is  widely used way to analyze the scattering mechanism, as for deeply virtual Compton scattering \cite{DGPR}. In the neutrino case, it reads:
\begin{eqnarray}
\label{cs}
&&\frac{d^4\sigma(\nu N\to l^- N'D)}{dx_B\, dQ^2\, dt\,  d\varphi}=\tilde\Gamma
\Bigl\{ \frac{1+ \sqrt{1-\varepsilon^2}}{2} \sigma_{- -}+\varepsilon\sigma_{00}
%-\varepsilon\cos(2\varphi)\sigma_{+-}
\\
&& +  \sqrt{\varepsilon}(\sqrt{1+\varepsilon}+\sqrt{1-\varepsilon} )(\cos\varphi\
{\rm Re}\sigma_{- 0} + \sin\varphi\
 {\rm Im}\sigma_{- 0} )\ \Bigr\},\nonumber
\end{eqnarray}
where the ``cross-sections'' $\sigma_{lm}=\epsilon^{* \mu}_l W_{\mu \nu} \epsilon^\nu_m$ are product of  amplitudes for the process $ W(\epsilon_l) N\to D N' $, averaged  (summed) over the initial (final) hadron polarizations.
The azimuthal angle $\varphi$ is defined in the  initial nucleon  rest frame as: 
\begin{equation}
sin ~\varphi = \frac {\vec q \cdot[(\vec q \times \vec p_D) \times (\vec q \times \vec k)]}{|\vec q||\vec q\times \vec p_D||
\vec q\times \vec k|}\,,
\end{equation}
while the final nucleon momentum lies in the $xz$ plane ($\Delta^y = 0$).

 We now calculate from $T_{L} $ and $T_{T} $ the  quantities $\sigma_{00}$, $\sigma_{--}$ and  $\sigma_{-0}$ which enter the differential cross section (\ref{cs}).
 The longitudinal and transverse cross sections $\sigma_{00}$  and $\sigma_{--}$, at zeroth order in $\Delta_T$, depend on longitudinal ${\cal H }_D, \tilde{\cal H }_D, \tilde{\cal E }_D$ and transverse   form factors through :
\begin{eqnarray}
\sigma_{00} &=&    \frac{C^2} {2 Q^2}\biggl\{8 (|{\mathcal{H}}^2_{D}| + |\tilde{\mathcal{H}}^2_{D}|)(1-\xi^2) + |\tilde{\mathcal{E}}^2_{D}|\frac{1+\xi^2}{1-\xi^2}\biggr\} \,,\\
\sigma_{--} &=&   \frac{4\xi^2 C^2 m_c^2}{Q^4} \biggl\{(1-\xi^2)|{\mathcal{H}_T^\phi}|^2  + \frac{\xi^2}{1-\xi^2} | {\bar {\mathcal{E}}_T^\phi}|^2
-2\xi \mathcal{R}e [ \mathcal{H}_T^\phi {\bar {\mathcal{E}}_T^{\phi *}}]\biggr\} . 
\end{eqnarray}
The  interference cross section   $\sigma_{-0}$ has its first non-vanishing contribution  linear  in $\Delta_T/m_N$ (with $\lambda = \tau^* = 1+i2$):
\begin{eqnarray}
&&\sigma_{-0} = \frac{- \xi \sqrt 2 C^2}{m_N}  \frac{ m_c}{Q^3} \,
\biggl\{-i \mathcal{H}_T^{*\phi} \tilde \mathcal {E}_D \xi(1+\xi)\epsilon^{pn\Delta  \lambda } 
 +   \mathcal{ H}_T^{*\phi} \Delta^\lambda [ -(1+\xi) \mathcal {E}_D]  \nonumber \\
&& +   \tilde \mathcal{ H}_T^{*\phi} \Delta^\lambda [2\mathcal {H}_D -\frac{2\xi^2}{1-\xi^2} \mathcal {E}_D]  +  \mathcal{E}_T ^{*\phi} \Delta^\lambda [(1-\xi^2) \mathcal {H}_D  - \xi^2\mathcal {E}_D] \\
&& +  {\bar {\mathcal{E}}_T^{\phi *}}[ \Delta^\lambda [(1+\xi)\mathcal {H}_D  +\xi \mathcal {E}_D]+i (1+\xi)\epsilon^{p n \Delta \lambda } \tilde \mathcal {H}_D  ]\ \biggr\}.\nonumber 
\end{eqnarray}
 The quantities  ${\cal R}e (\sigma_{- 0}) $ and ${\cal I}m (\sigma_{- 0})$  are directly related to the observables $<cos \varphi>$ and $<sin \varphi>$ through
     \begin{eqnarray}
  <cos ~\varphi>&=&\frac{\int cos ~\varphi ~d\varphi ~d^4\sigma}{\int d\varphi ~d^4\sigma}= K_\epsilon\, \frac{{\cal R}e \sigma_{- 0}}{\sigma_{0 0}}  \,, \nonumber \\
   <sin~ \varphi>&=&K_\epsilon \frac{{\cal I}m \sigma_{- 0}}{\sigma_{0 0}}  \,,
   \end{eqnarray} 
   with $K_\epsilon =\frac{\sqrt{1+\varepsilon}+\sqrt{1-\varepsilon} }{2 \sqrt{\epsilon} }$ and where we consistently neglected the $O(\frac{m_c^2}{Q^2})$ contribution of $\sigma_{--}$ in the denominator. 
In the legitimate limit of small $\alpha$, the dependence on the heavy meson DA effectively  disappears in the  ratios of the r.h.s. of Eq. (11). The complete formula  is quite lengthy but a simple approximate result reads:
  \begin{eqnarray}
&& \hspace{-1cm}<cos\varphi> \approx \frac{K{\cal R}e[{\mathcal{H}}_{D} (2{\tilde \mathcal{ H}}_{T}^\phi + {\mathcal{E}}_{T}^\phi + {\bar {\mathcal{E}}_T^\phi})^*- \mathcal {E}_D {\mathcal{H}}_{T}^{\phi *}]} {8|{\mathcal{H}}^2_{D}| + |\tilde{\mathcal{E}}^2_{D}|} \,, \nonumber \\
 && \hspace{-1cm}<sin\varphi> \approx  \frac{K{\cal I}m[{\mathcal{H}}_{D} (2 {\tilde \mathcal{ H}}_{T}^\phi + {\mathcal{E}}_{T} ^\phi+ {\bar {\mathcal{E}}_T^\phi})^*- \mathcal {E}_D {\mathcal{H}}_{T}^{\phi *} ]}{8 |{\mathcal{H}}^2_{D}| + |\tilde{\mathcal{E}}^2_{D}|} \, ,\nonumber \\
K&=& -\frac{\sqrt{1+\varepsilon}+\sqrt{1-\varepsilon} }{2 \sqrt{\epsilon} } ~\frac{ 2\sqrt 2 \xi  m_c}{Q } \, \frac{\Delta_T}{m_N} \,.
    \end{eqnarray} 
\section{Conclusion.}
We thus have defined a new way to get access to the transversity chiral-odd generalized parton distributions, the knowledge of which would shed  light on the tomography of the quark structure of the nucleons.  Planned high energy neutrino facilities \cite{NOVA}, which have their scientific program oriented toward the understanding of neutrino oscillations  or elusive sterile neutrinos, may thus allow some important progress in the realm of hadronic physics.

\vspace{.1cm}
\noindent
 This work is partly supported by the Polish Grant NCN No DEC-2011/01/B/ST2/03915 and  the French grant ANR PARTONS (Grant No.ANR-12-MONU-0008-01).

\end{document}